\documentclass{emulateapj}
\usepackage{graphicx,natbib}

\begin{document}

\title{Numerical simulations of shock wave-driven jets}

\author{L. Heggland}
\affil{Institute of Theoretical Astrophysics, University of Oslo,
  P.O. Box 1029, Blindern, N-0315 Oslo, Norway}
\email{lars.heggland@astro.uio.no}

\author{B. De Pontieu}
\affil{Lockheed Martin Solar and Astrophysics Laboratory, 3251 Hanover Street,
  Org. ADBS, Building 252, Palo Alto, CA 94304}

\author{V. H. Hansteen\altaffilmark{1}}
\affil{Institute of Theoretical Astrophysics, University of Oslo,
  P.O. Box 1029, Blindern, N-0315 Oslo, Norway}

\altaffiltext{1}{Also at Center of Mathematics for Applications,
  University of Oslo, P.O. Box 1053, Blindern, N-0316 Oslo, Norway}

\begin{abstract}

We present the results of numerical simulations of shock wave-driven
jets in the solar atmosphere. The dependence of observable quantities like
maximum velocity and deceleration on parameters such as the period and
amplitude of initial disturbances and the inclination of the magnetic field
is investigated. Our simulations show excellent agreement with observations,
and shed new light on the correlation between velocity and deceleration
and on the regional differences found in observations.

\end{abstract}


\section{Introduction}

In the solar chromosphere, jet-like dynamic features are found in several
regions. At the quiet sun limb, we find spicules, chromospheric protrusions
that reach heights of 5-9 Mm and last for 3-15 minutes, reaching velocities
of 10-30 km s$^{-1}$ \citep{Beckers1968}. Many models have been proposed to
explain
their formation \citep{Sterling2000}, but observational and interpretational
difficulties have made the models hard to constrain.

On the quiet sun disk, we observe dark mottles, which appear to have many
similarities to spicules. There has been some controversy over the
relationship between spicules and mottles, with
\citet{Grossmann-Doerth+Schmidt1992}
concluding that they are not counterparts, while other authors have
argued that the similarities are striking
\citep{Tsiropoula+etal1994,Suematsu+etal1995,Christopoulou+etal2001}.

A third group which has come under study recently
\citep{DePontieu+etal2004,DePontieu+etal2007,Hansteen+etal2006} 
are dynamic fibrils, a subset of jets frequently 
found in the vicinity of active region plage. These jets are shorter than
spicules in both length and duration, reaching heights of 1-4 Mm and lasting 
3-6 minutes, and frequently show both periodicity and internal structure
varying on shorter timescales.

Recent observational evidence \citep{Hansteen+etal2006,DePontieu+etal2007}
has suggested that dynamic fibrils are driven by magnetoacoustic shocks,
which originate in the convection zone and photosphere and, although
usually evanescent in the chromosphere, may be able to leak into the
upper layers in inclined or heated flux tubes. The data of
\citet{Rouppe+etal2007} indicate that the same mechanism may also be the 
driving force behind at least a subset of quiet sun mottles.

We perform numerical simulations of such shock wave-driven jets (henceforth 
called fibrils), and analyse the data in a similar fashion to 
\citet{Hansteen+etal2006}, \citet{DePontieu+etal2007} and 
\citet{Rouppe+etal2007}. With a more idealised and controlled environment, 
we can study
the effect of parameters such as the period and amplitude of the piston
driver and the inclination of the magnetic field. We examine
the correlation between the maximum velocity of a fibril and its deceleration,
and also see if we can confirm the suggestion 
\citep{Michalitsanos1973,Bel+Leroy1977,Suematsu1990,DePontieu+etal2004,
DePontieu+etal2005,DePontieu+etal2007,Hansteen+etal2006} that inclined 
magnetic fields can allow normally evanescent
long-period waves to propagate into the upper chromosphere and corona.
Signs of such waves have recently been observed \citep{DePontieu+etal2005,
McIntosh+Jefferies2006,Jefferies+etal2006}.

\section{Simulations}

For our simulations, we use a simple one-dimensional model of the upper
solar atmosphere, with a monochromatic piston driver at the lower
chromospheric boundary for creating acoustic waves. As these waves
travel upwards, they gain in amplitude because of the decreasing density
of the medium and steepen into shocks (assuming a large enough initial
amplitude), which hit the transition region and thereby push the corona
upwards.

\begin{figure}[ht]
 \begin{center}
  \plotone{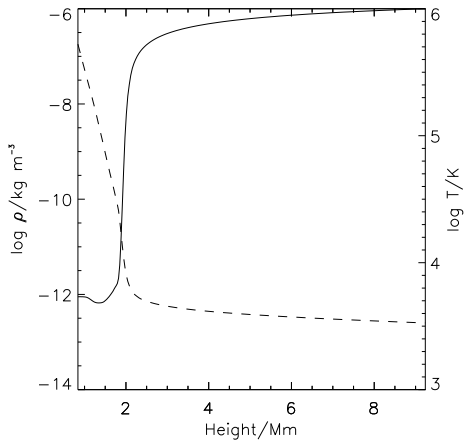}
 \end{center}
 \caption{Initial density (dashed, left axis) and temperature (solid,
          right axis) structure of the model.}
 \label{fig:DensTemp}
\end{figure}

Fig.~\ref{fig:DensTemp} shows the initial density and temperature profiles
of our model atmosphere. The model extends around 8.5~Mm in height, starting
0.85~Mm above the photosphere, with the
transition region about 1~Mm above the lower boundary, the chromosphere
below and the corona above it. The temperature at the upper boundary is
maintained at 1~MK while heat conduction and radiation set the temperature
in the rest of the domain.

In computing the radiative losses we include a radiative loss term
that accounts for the energy loss due to collisional excitation of the
various ions comprising the plasma.  We have included the elements
hydrogen, carbon, oxygen, neon, and iron, as well as thermal bremsstrahlung, 
using the ionization and recombination rates
given by \citet{Arnaud+Rothenflug1985} and \citet{Shull1982} and using the 
collisional excitation rates found through the
HAO-DIAPER atom data package \citep{Judge+Meisner1994}.
The metals are treated by assuming ionization
equilibrium and then deriving an \emph{a priori} radiative loss curve
as a function of electron temperature.  While one should ideally solve the
equation of radiative transport in order to calculate the radiative
losses from hydrogen, comparison with models where this has been done 
indicate that the errors incurred by assuming effectively thin losses
in the Ly$\alpha$ line are not significant to the processes studied
in this paper. In order to avoid too large radiative losses in the
lower atmosphere the radiative loss term is multiplied with
$\exp(-\tau)$, where the ``optical depth'' $\tau$ is set proportional
to the gas pressure. 

Thermal conduction is given by $\kappa_0T^{5/2}{dT/ds}$, which carries
energy from the corona to the upper chromosphere and determines the
temperature structure of the transition region. The numerical model is
set up so that conduction is carried along the magnetic field. The
energy equation is solved by operator splitting with thermal 
conduction solved implicitly using a multi-grid method as described by
\cite{Hansteen2005}. 

Although it might have been ideal to include the solar atmosphere from
the photosphere to the corona in our model, we have chosen to restrict
ourselves to the magnetically dominated upper regions, from
the chromosphere and upwards. This is because, with an inclined magnetic
field, the $\beta \approx 1$ layer, where $\beta$ is the ratio between the
thermal and magnetic pressure, is an area of extensive conversion
between the so called fast and slow modes
\citep{Heggland2003,Bogdan+etal2003}.
We are here primarily interested in the field-aligned acoustic
(upper-atmosphere slow mode) waves, and not the magnetosonic fast mode
waves that can propagate across field lines. The reason is that, in a
real multidimensional atmosphere, the fast mode waves will be refracted
into areas of low Alfv\'{e}n speed \citep[e.g.][]{Osterbrock1961}, 
and thus have difficulty
propagating into the higher layers of the atmosphere. In a 1D model, they
are forced to propagate vertically, and will then reach the upper layers
with far greater energy than is realistic. For this reason, we get cleaner
and arguably more realistic results by placing our driving piston in the
low-$\beta$ region of the (upper) chromosphere, and restricting our domain
to the areas above that.

Of course, $\beta$ could be reduced to any value we like simply by increasing
the strength of the magnetic field. We have chosen a field strength of
$6.0 \times 10^{-3}$ T (60 gauss), representing a reasonable value for
chromospheric field outside of, but not too far from, active regions.
A stronger field would allow us to include deeper layers of the atmosphere,
but at the cost of shorter timesteps because of the increased Alfv\'{e}n
speed. We believe that including deeper layers would be unlikely to change
our results significantly, especially in light of the other simplifications
inherent in a 1D model. In 1D, we are primarily interested in studying
fundamental effects, and further refinements should rather be considered
in future multi-dimensional investigations.

The code we use is a version of the code described in 
\citet{Hansteen2005}. (See also Galsgaard \& Nordlund's description of an 
earlier version of the code at \texttt{http://www.astro.ku.dk/\~{}kg}).
This code allows us
to use 3D variables in a 1D domain, thus allowing for transverse components
of the velocity and magnetic field. We exploit this by using the same
background atmosphere (density/temperature) in all cases but tilting the
magnetic field away from the vertical.

With plasma movement thereby mostly constrained to being along the inclined
magnetic field, but having a vertical computational domain, the simulated
data are basically projections of the field-aligned motion onto the
vertical. As a result, propagation speeds appear lower when the field
is highly inclined --- at $60^\circ$, the waves have to travel twice as
far along the field to reach the same height. If desired, the vertical 
velocities can be recalculated as field-aligned by multiplication with 
a simple 1/cos $\theta$ factor, $\theta$ being the inclination. 

Using a 1D model makes our analysis simpler, while keeping the important
physics. The main phenomena we exclude are wave refraction, which (as
mentioned) is only important for the fast mode, and the curvature and
expansion of the magnetic field.

In our simulations, the 60 gauss field is constant and homogeneous.
We vary the inclination, $\theta$, from
$0^\circ$ (vertical) via $30^\circ$ and $45^\circ$ to $60^\circ$. We
use piston periods of 180, 240, 300 and 360~s, and initial amplitudes
of 200, 500, 800 and 1100~m~s$^{-1}$ at the lower boundary. The driving
is sinusoidal and the piston is active throughout the simulations.
The piston movement is along the magnetic field rather than vertical, in
order to further suppress unwanted fast modes.

The piston generates a train of waves that quickly (especially at greater
initial amplitudes) steepen into shocks. As these shocks hit the transition
region, they give a large impulsive acceleration (a ``kick'') to the plasma
there as the shock front passes, then a more gentle ``push'' for a while
in the receding phase of the shock. Some of the wave energy passes through
and enters the corona, but quite a lot is reflected down again ---
\citet{Heggland2003} estimates a reflection coefficient of about 0.70 for
low frequency linear waves, though shock waves may behave slightly
differently.

\begin{figure}[p]
 \begin{center}
  \plotone{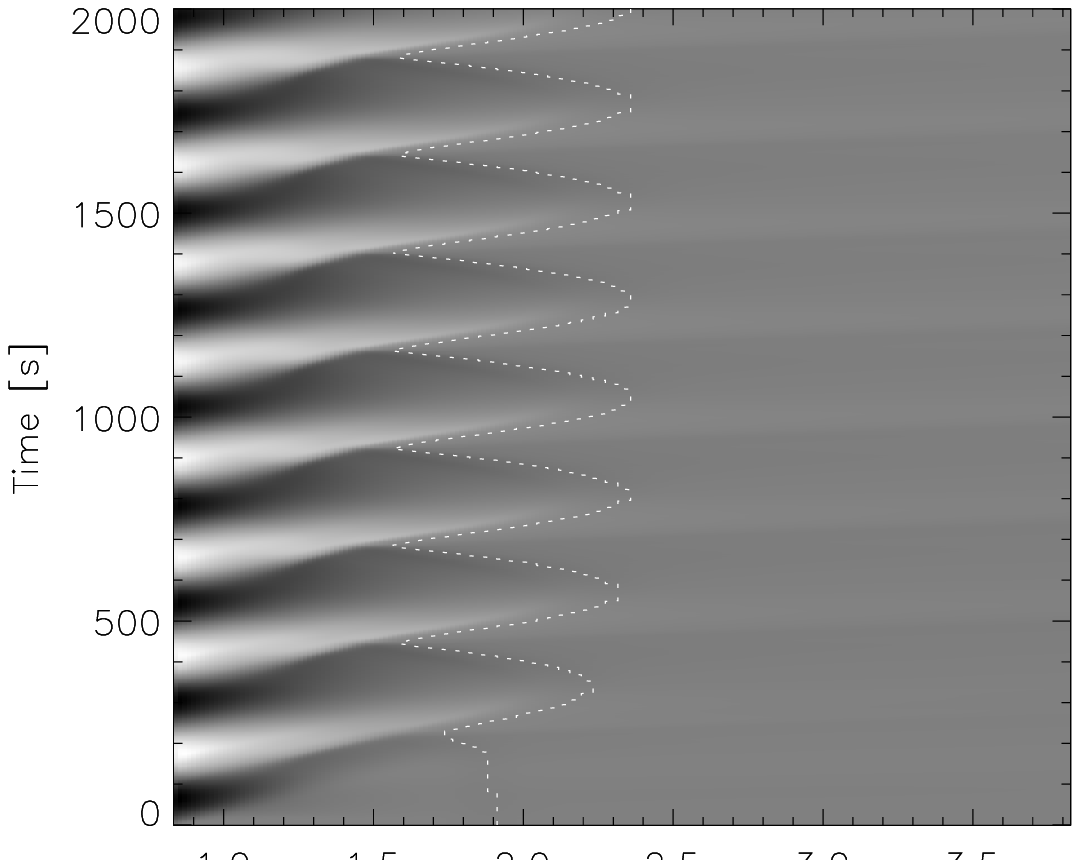}
 \end{center}
 \caption{The vertical energy flux density of the waves in
          an example case ($45^\circ$ inclination,
          240~s driver period and 1100~m s$^{-1}$ initial amplitude). 
          We have plotted the square root of the actual energy flux density
          in order to preserve the sign of the velocity --- white represents
          upward velocity and black downward. Most of the energy
          ends up in standing waves in the chromosphere, while only a
          small fraction (almost invisible in this plot) reaches the corona.
          The dotted line is at the height where the temperature is
          100~000~K, i.e. the lower transition region.}
 \label{fig:eflux}
\end{figure}

In Fig.~\ref{fig:eflux}, we have plotted a representation of the vertical 
energy flux for an
example case as a function of height and time. Reflection
from the transition region and re-reflection from the closed lower boundary
lead to the formation of standing waves in the chromosphere, which makes
reliable estimation of the upward energy flux difficult; however, crude
estimates comparing the input energy from the piston with the energy flux
reaching the upper boundary give a transmission coefficient of 10\% or less,
the rest being reflected or dissipated.

\begin{figure}[p]
 \begin{center}
  \plotone{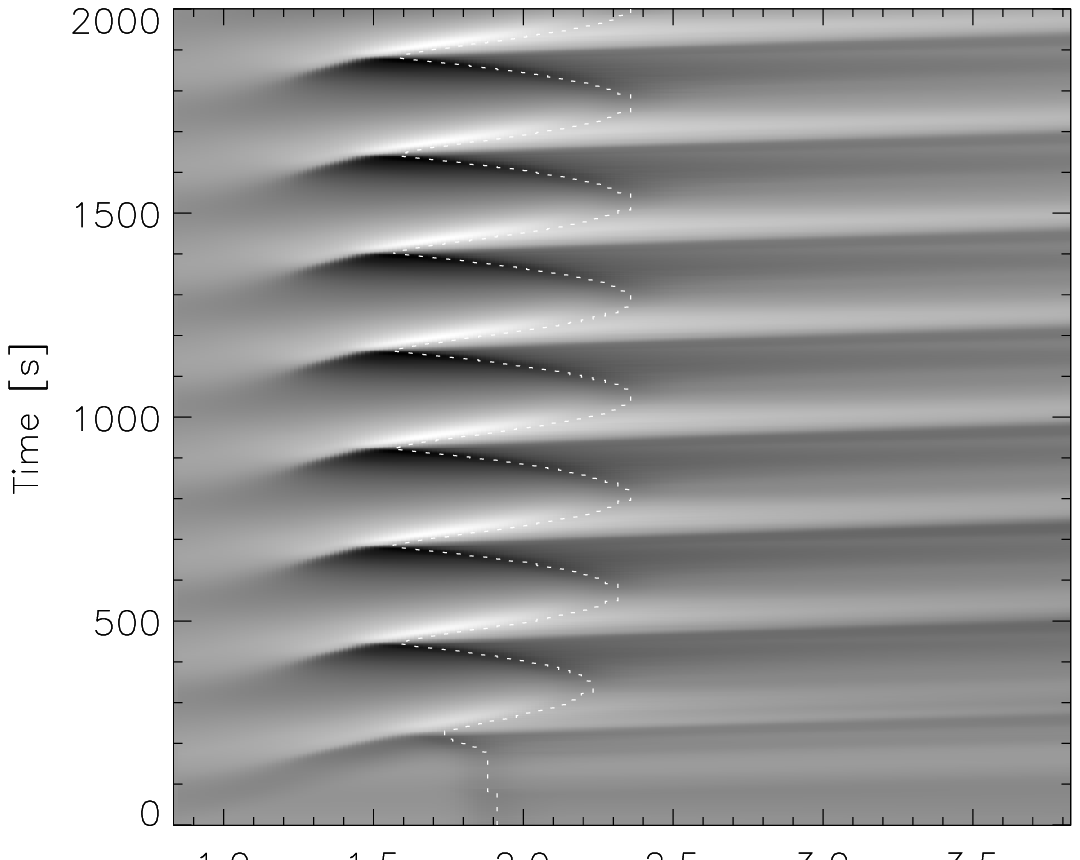}
 \end{center}
 \caption{The vertical velocity taken directly from the simulation data in the
          same case as Fig.~\ref{fig:eflux}. In
          addition to the main fronts and the parabolic shapes, we see
          an initial (quite weak) fast mode front and some
          additional fronts that seem to be generated at the tops of the
          fibrils. Again, the dotted line marks where T=100~000~K.}
 \label{fig:uzdirect}
\end{figure}

A plot of the simulated vertical velocity as a function of height 
and time for an example case is found in Fig.~\ref{fig:uzdirect}. In this
case we hardly see the standing waves because of the large amplitudes
the shock waves reach when they enter the transition region where the
density falls off rapidly with height.

\section{Analysis}

\begin{figure}[p]
 \begin{center}
  \plotone{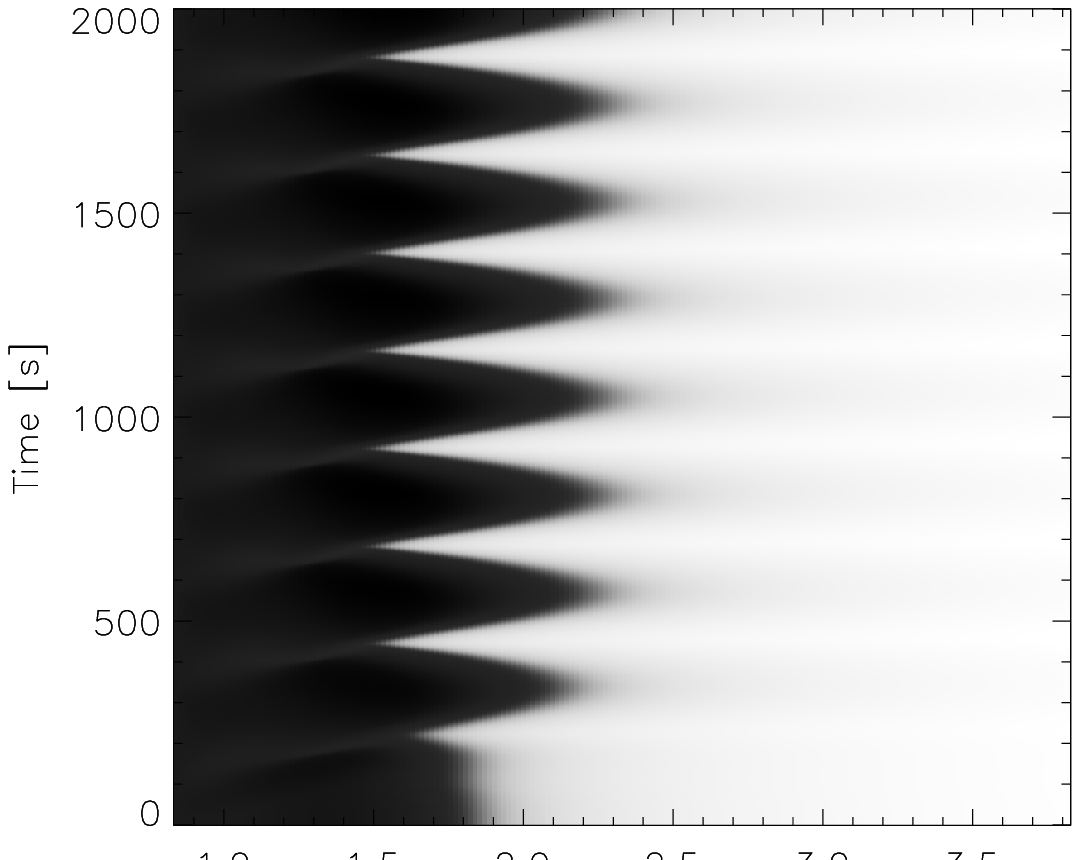}
 \end{center}
 \caption{The gas temperature as a function of time and vertical position in
          the same case as for Fig.~\ref{fig:uzdirect}. This way of plotting
          very clearly shows the movement of the transition region which is our
          primary interest.}
 \label{fig:tgvszt}
\end{figure}

In their observational study, \citet{Hansteen+etal2006} and
\citet{DePontieu+etal2007} find that fibrils,
which they observe in H$\alpha$, follow parabolic paths in distance-time 
diagrams. These parabolic paths are found in our simulations as well,
cf. Fig.~\ref{fig:uzdirect}.
They can be visualized in a way more comparable to
\citeauthor{DePontieu+etal2007} if we instead show the \emph{temperature} in a
distance-time plot, as in Fig.~\ref{fig:tgvszt}. We clearly see the
parabolic paths traced by the transition region as it is periodically
pushed up by the passing shock waves.
H$\alpha$ radiation is primarily generated in the hot upper layers of the
chromosphere just below the transition region, and observations in that
band will therefore show very similar movement.

A look at Fig.~\ref{fig:uzdirect} tells us that, although the parabolic
shapes are clearly seen, there are also other features present. Some
fast mode waves remain --- one can be seen faintly at the beginning of the
simulation, before the first slow shock front ---
and some additional slow mode waves are apparently generated in the
shock fronts through nonlinear processes. Although these other waves
can have significant
amplitudes, their effect on the movement of the transition region is
far less than that of the slow mode shock waves. In addition, fast 
modes would tend to refract downwards in a real atmosphere, as discussed in
the previous section.

\begin{figure}
 \begin{center}
  \plotone{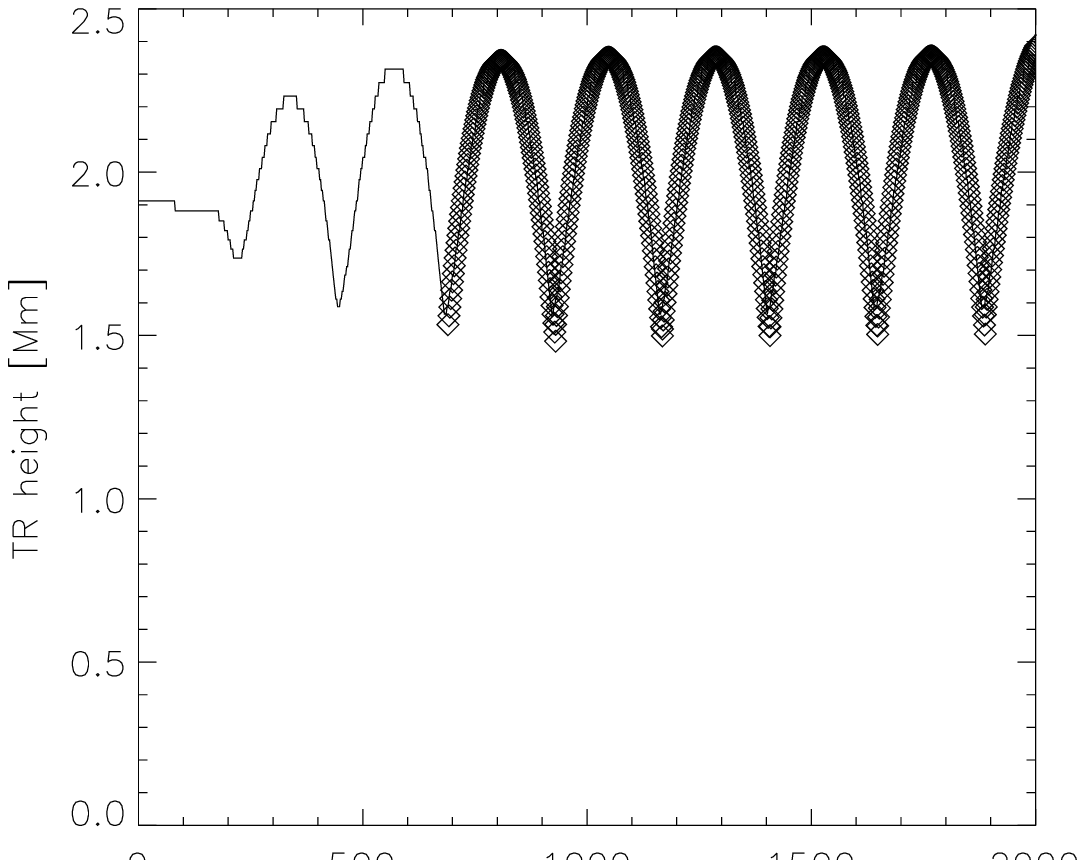}
 \end{center}
 \caption{An example of the parabolic fitting process, for the same case as
          Figs.~\ref{fig:uzdirect} and~\ref{fig:tgvszt}. The transition
          region height is defined as the height where the temperature
          is 100~000~K.}
 \label{fig:parabolas}
\end{figure}

The picture is clearer when looking at the temperature plot. Therefore,
it seems reasonable to use the movement of the transition region as a proxy
for the movement of the layers where the main H$\alpha$ radiation takes
place, make parabolic fits of the position, and use derivatives of these
to determine velocities and decelerations. This is similar to the method used 
by \citet{DePontieu+etal2007}, and makes comparisons easier. Furthermore,
it means that the calculated decelerations will be constants. An example
of the parabolic fits is shown in Fig.~\ref{fig:parabolas}.

\begin{figure}[p]
 \begin{center}
  \plotone{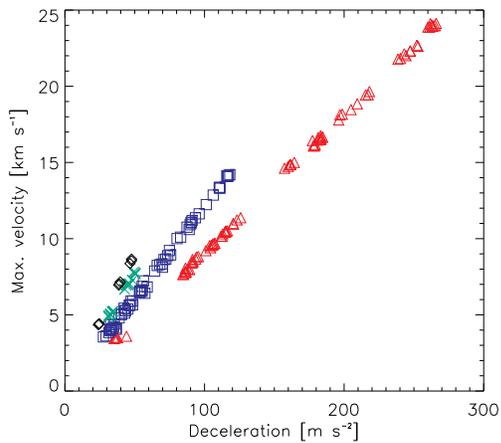}
 \end{center}
 \caption{Scatter plot of the maximum (upward or downward) velocities of
          fibrils vs. their decelerations, showing clear linear correlations.
          Red triangles correspond to a period of 180~s, blue squares to
          240~s, green crosses to 300~s and black diamonds to 360~s. The
          slopes are clearly different as the period increases.}
 \label{fig:decvelcorr}
\end{figure}

\begin{figure}[p]
 \begin{center}
  \plotone{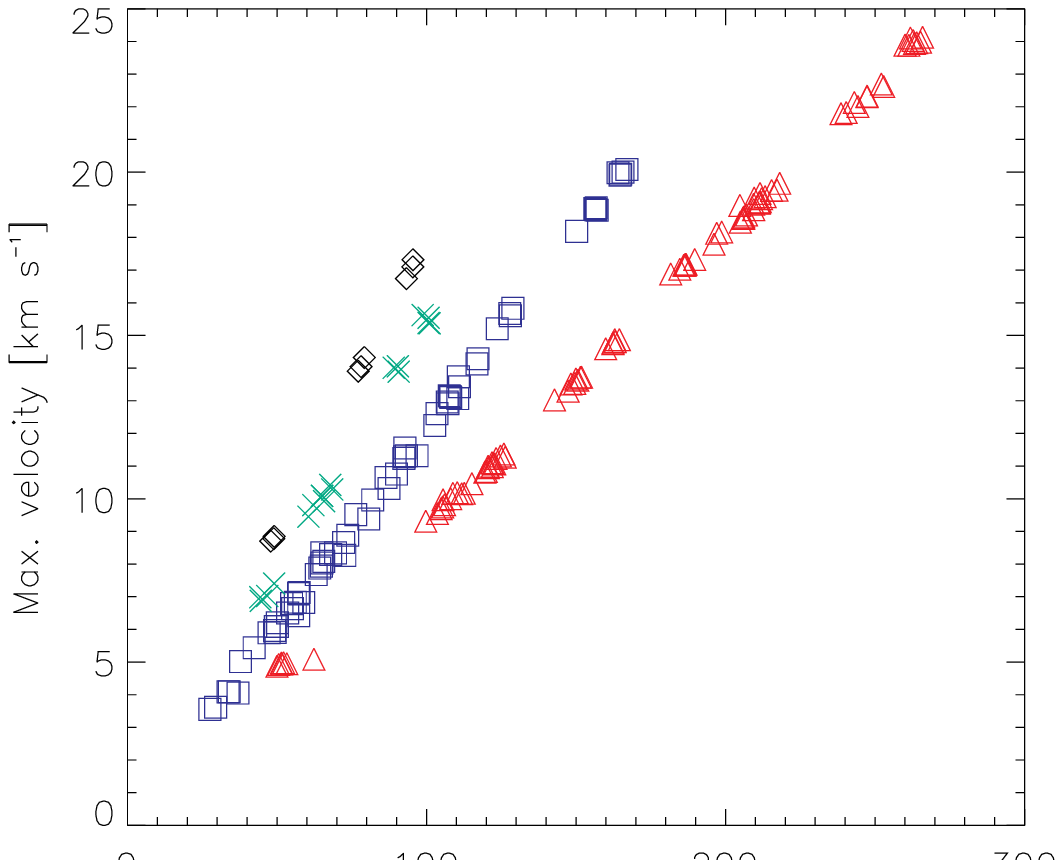}
 \end{center}
 \caption{As Fig.~\ref{fig:decvelcorr}, but adjusted for projection effects.}
 \label{fig:decvelcorradj}
\end{figure}

Having determined the decelerations and velocities of each fibril from our 
64 simulated
cases (not all of which lead to noticeable movement of the transition region),
a total of 190 fibrils, we can make scatterplots in the same way as 
\citeauthor{DePontieu+etal2007} The
results are shown in Figs.~\ref{fig:decvelcorr} and~\ref{fig:decvelcorradj}
--- the latter is corrected for projection effects through multiplication by
1/cos $\theta$ when the field is inclined. We see obvious linear
correlations between deceleration and maximum velocity, but the slopes
are clearly different in the 180~s and 240~s cases. Although the number of
points for the 300~s and 360~s cases is small, they too appear to show the
same trend, giving progressively lower decelerations for a given maximum
velocity. The data of \citeauthor{DePontieu+etal2007} are much more scattered, 
both because of the greater difficulty in analysing multi-dimensional
observational data and because of uncertainty in the relative orientations
of the fibrils' movement and the line of sight, but they also find a more
or less linear correlation and signs of a period-dependent slope ---
again giving smaller decelerations at greater periods.

\begin{figure}
 \begin{center}
  \plotone{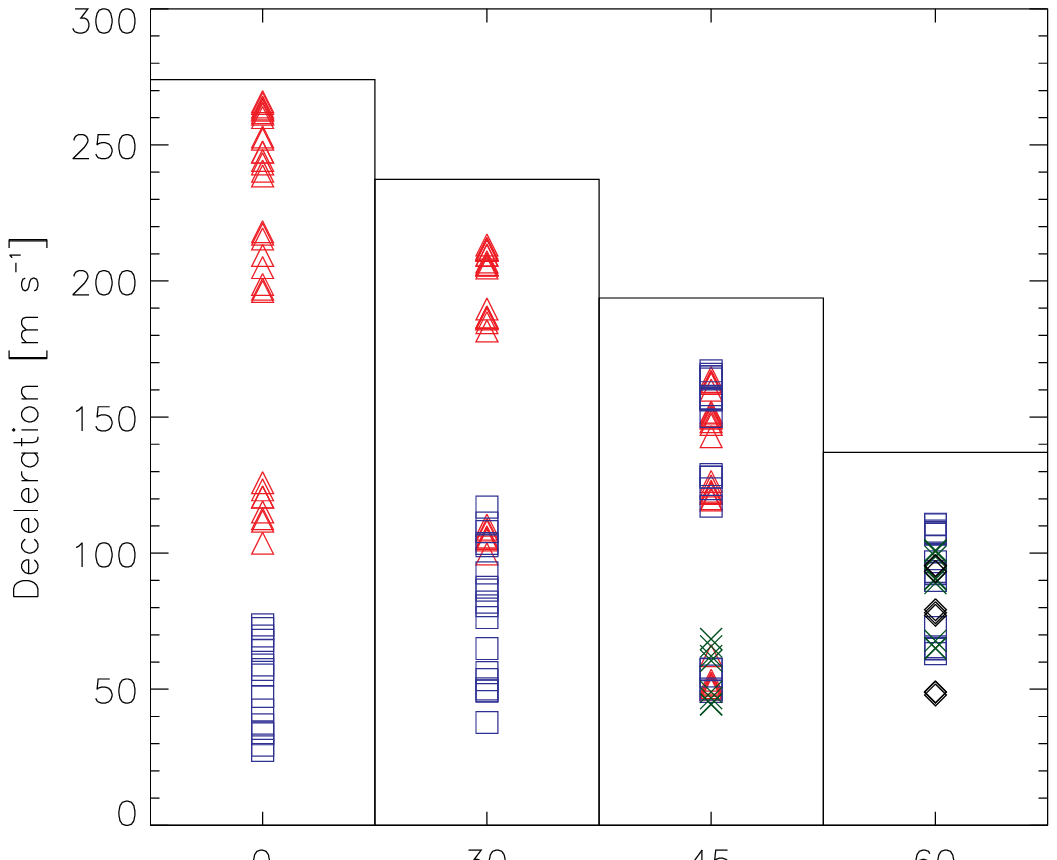}
 \end{center}
 \caption{Distribution of decelerations with magnetic field inclination.
          The data symbols are the same as in Fig.~\ref{fig:decvelcorr}.
          All decelerations are lower than the projected gravity (solid
          columns), often very much lower. Since we have already compensated
          for projection effects, this distribution is extremely hard to
          reconcile with a ballistic theory.}
 \label{fig:dechisto}
\end{figure}

A question of interest is how big a role gravity plays in setting the
deceleration. One natural instinct when seeing the parabolic paths is to
assume that we are observing a ballistic flight under the influence of
solar gravity, with differences being accounted for by inclination and
line of sight effects. In observations, the angle between the fibril
movement and the line of sight is often difficult to determine exactly,
but the many low decelerations reported would require extremely steep
inclinations for the fibrils \citep{Suematsu+etal1995,DePontieu+etal2007}.

In looking at our simulation data, we have no line of sight uncertainty,
and compensating for the inclination of the fibrils is trivial. We can
therefore easily calculate the distribution of decelerations with inclination
angle. This distribution is shown in Fig.~\ref{fig:dechisto}. The black
histogram shows the field-aligned component of the solar gravitational 
acceleration, while the data points are shown with the same shapes and
colour scheme as used in Fig.~\ref{fig:decvelcorr}. We see that all our
observed decelerations are smaller than the projected gravity, and often
significantly smaller. Furthermore, they are not particularly clustered,
pretty much the whole range of decelerations from about 50~m~s$^{-1}$ to
gravity being represented at all angles. With all projection effects
already accounted for, this distribution is extremely hard to reconcile
with a ballistic model, and indicates that we should look not to gravity
but elsewhere in trying to explain it.

\begin{figure}
 \begin{center}
  \plotone{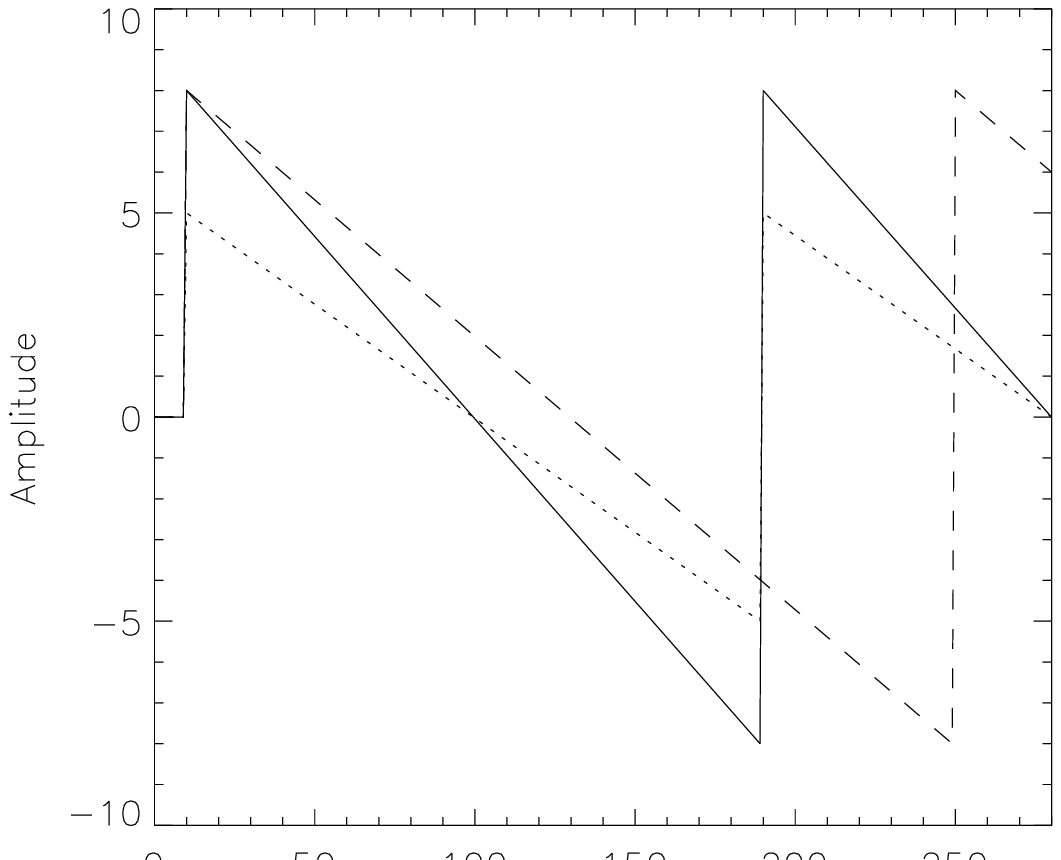}
 \end{center}
 \caption{An N-wave with lower amplitude (dotted line) will have a less
          steep descending slope (lower deceleration) than one with higher
          amplitude (solid line) at a given period. Similarly, one with a
          longer period (dashed line) will have a
          lower deceleration than one with a shorter period at a given
          amplitude.}
 \label{fig:waveforms}
\end{figure}

One alternative is to look at shock wave physics and pressure forces.
The period of a wave train should remain constant during its
propagation, and in that case, a fully developed shock wave (an N-wave)
has a given time during which the amplitude must move from its highest
to its lowest value. If the maximum amplitude --- approximately the same
magnitude as the maximum velocity of the fibril --- increases, the deceleration
must then be greater per unit time. Similarly, if the period is longer,
the wave has more time between maximum and minimum amplitude, and the
slope becomes less steep, meaning the deceleration is lower for a given
amplitude. This is illustrated in Fig.~\ref{fig:waveforms}.

\begin{figure}
 \begin{center}
  \plotone{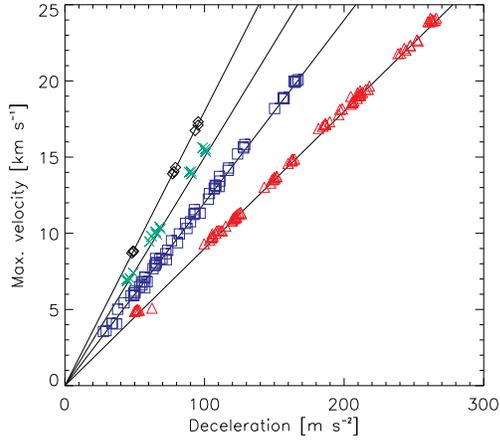}
 \end{center}
 \caption{As Fig.~\ref{fig:decvelcorradj}, but including the theoretical
          values from the shock deceleration hypothesis as solid lines.
          There is a near-perfect fit with the data, giving strong support
          to the hypothesis.}
  \label{fig:decvelcorrtheo}
\end{figure}

\begin{figure}
 \begin{center}
  \plotone{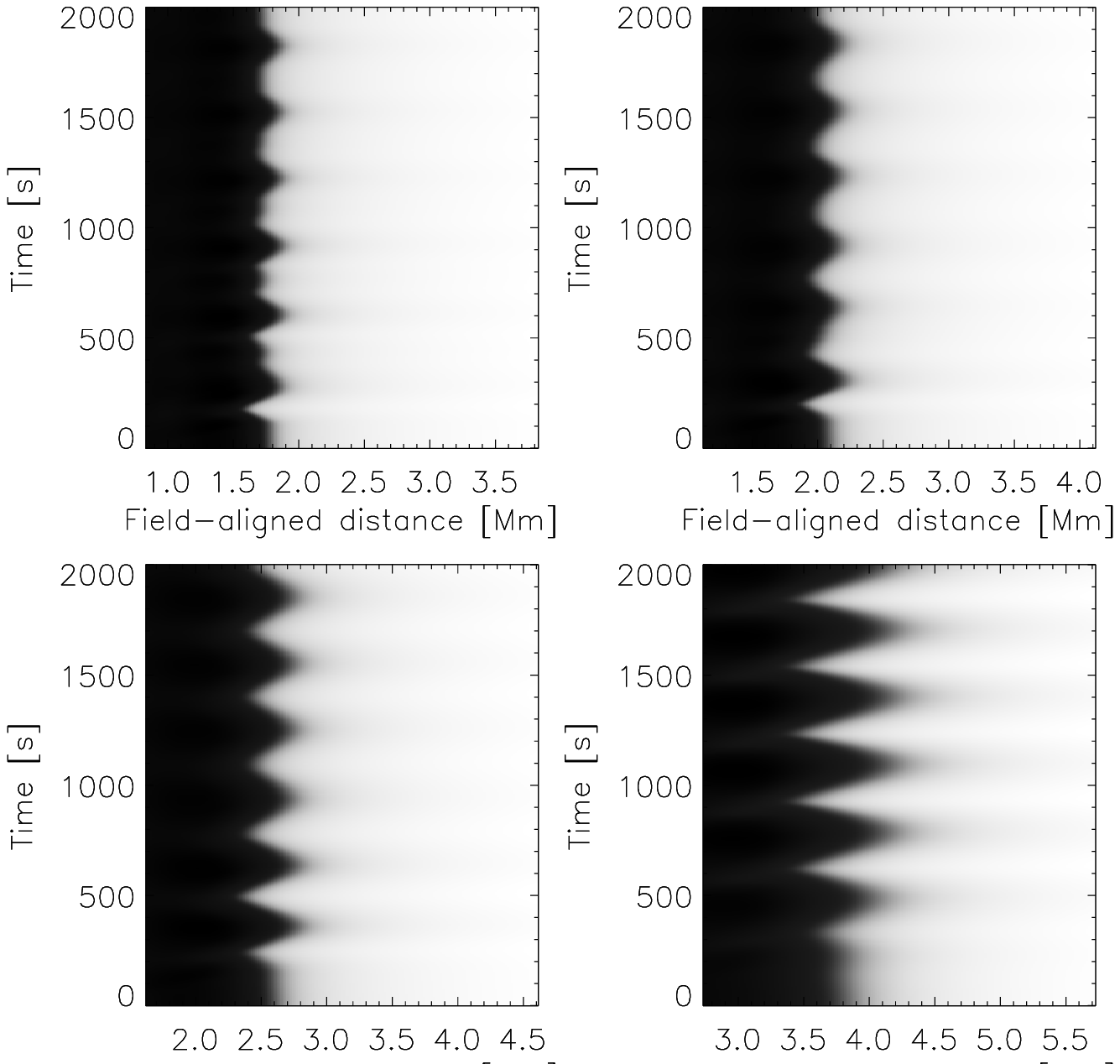}
 \end{center}
 \caption{Temperature plots showing the movement of the transition region
          with time. The period is 300~s; the piston amplitude is kept
          constant at 500~m s$^{-1}$, while
          the inclination of the magnetic field varies: $0^\circ$ (top left),
          $30^\circ$ (top right), $45^\circ$ (bottom left), $60^\circ$
          (bottom right). As the inclination increases, the movement of the
          transition region becomes more regular and also gains in amplitude.
          The horizontal axis is distance along the field rather than height,
          in order to show the true length of the fibrils. The left boundary
          is at a different coordinate in each panel, so that the unperturbed
          transition region appears in the same position, but all panels
          are at the same scale.}
 \label{fig:incline}
\end{figure}

The shock deceleration hypothesis can thus explain both the linear correlation
between deceleration and maximum velocity and the variation in its slope
with wave period. Indeed, we can calculate the decelerations expected from
the theory as a function of period and maximum velocity/amplitude and
compare them with the values from our simulations. The formula is simply
\begin{equation}
  d = \frac{v_{max}}{P / 2},
\end{equation}
where $d$ is the deceleration, $v_{max}$ is the maximum velocity and $P$ 
the period of the wave. As we see in
Fig.~\ref{fig:decvelcorrtheo}, there is near-perfect correspondence between
the theoretical and simulated values, giving
strong support to this explanation. 

Although all the fibrils we study in this paper have decelerations below
projected gravity, the shock wave deceleration model can also explain
decelerations greater than gravity. Indeed, in preliminary experiments with 
even stronger
piston driving we have produced decelerations that are slightly greater
than gravity. Such decelerations have also recently been reported in
observations of quiet sun mottles \citep{Rouppe+etal2007}.

One difference between our results and those of \citet{DePontieu+etal2007}
is that they find many more fast fibrils than we do. Significant numbers 
have maximum velocities up to 25-30~km s$^{-1}$, whereas
we only find such values for short-period waves propagating vertically.
In fact, there are few fibrils with greater than 20~km s$^{-1}$ maximum
velocity in our simulations.
It is also worth noting that the data of \citeauthor{DePontieu+etal2007} are 
not corrected for projection effects. We will touch on a possible explanation
for this discrepancy later in this section.

We also find fibrils with lower maximum velocity than the lowest found in
the observations. This is probably because the fibrils are quite easy
to find and isolate in our simulation data, while superposition effects
and background noise make small disturbances hard to detect in observations.
These fibrils show a clear correspondence with the wave movement in the
simulations, so they are not an effect of background noise there.

We now turn our attention to the question of whether the inclination of
the field helps long-period waves propagate and reach coronal heights.
\citet{Michalitsanos1973} and \citet{Bel+Leroy1977} made early investigations
into this phenomenon. The idea is that if
the plasma is confined to moving along the magnetic field, it will also
be subjected to a reduced effective gravity $g \cos \theta$, where $\theta$ 
is the inclination of the magnetic field. With the acoustic cutoff period
given as
\begin{equation}
  P_{ac} = \frac{4 \pi c_s}{\gamma g}, \label{eq:cutoff}
\end{equation}
where $\gamma$ is the ratio of specific heats and $c_s$ the sound speed,
a reduced effective gravity also leads to a higher cutoff period, thus
allowing longer-period waves to propagate. More recently,
\citet{Suematsu1990} and \citet{DePontieu+etal2004,
DePontieu+etal2005} have proposed that this mechanism may let solar
p-modes, with a period of around 5~minutes, propagate through the
chromosphere and provide the driving force for spicules and fibrils.
These waves are usually evanescent, having periods above the acoustic cutoff.
In our model atmosphere, the
cutoff period at the lower boundary is 213~s at 0$^\circ$ inclination, 
245~s at 30$^\circ$, 301~s at 45$^\circ$, and 425~s at 60$^\circ$. The
real solar chromosphere should have values close to these.

Note that the reduced effective gravity hypothesis only works if the plasma
is in fact confined to moving along the magnetic field, that is, in 
low-$\beta$ plasma. In our model this applies everywhere, but on
the actual sun, where the main source of wave energy is the convection
region and photosphere, we would only expect long-period waves to be able
to propagate upwards from the photosphere with significant energy in areas
where the magnetic field
is strong enough to dominate over thermal pressure forces even at
photospheric depths. This applies mainly in the vicinity of sunspots, active
regions and network flux concentrations.

In Fig.~\ref{fig:incline} we have plotted the movement of the transition
region with time for four different inclinations, with the other parameters
--- piston period and amplitude --- constant.
The period is 300~s, and waves with this period ought to be evanescent
in a vertical field, being above the cutoff period of 213~s.

With a vertical field (top left), there is some tunneling of wave energy,
but the movement is slight and somewhat irregular. As the inclination of
the field increases to $30^\circ$ (top right), $45^\circ$ (bottom left) and
$60^\circ$ (bottom right), there is a clear progression towards more regular 
parabolic movement, and the amplitude also increases markedly.
These results clearly support the hypothesis of \citet{Suematsu1990} and
\citet{DePontieu+etal2004,DePontieu+etal2005}. They also support the claim
by \citet{DePontieu+etal2007} that the regional differences in behaviour
they observe are at least partially caused by the inclination of the
magnetic field, allowing longer period waves (with corresponding lower
fibril decelerations) to propagate in regions of inclined field.

\begin{deluxetable}{crrrr}
\tablecolumns{5}
\tablewidth{0pt}
\tablecaption{Maximum velocities (km s$^{-1}$) \label{tbl:velocities}}
\tablehead{
  \multicolumn{5}{c}{0$^\circ$ inclination} \\
  \tableline \\[-5pt]
  \colhead{} & \multicolumn{4}{c}{Driver amplitude (km s$^{-1}$)} \\
  \colhead{Period (s)} & \colhead{\phn 0.2} & \colhead{\phn 0.5} & 
                         \colhead{\phn 0.8} & \colhead{\phn 1.1}}
\startdata
180 &    10.6 &    18.7 &    22.2 &    24.0 \\
240 & \nodata & \nodata &     6.2 &     8.4 \\
300 & \nodata & \nodata & \nodata & \nodata \\
360 & \nodata & \nodata & \nodata & \nodata \\
\cutinhead{30$^\circ$ inclination}
180 &     9.9 &    17.1 &    18.7 &    19.2 \\
240 & \nodata &     6.1 &     9.7 &    12.9 \\
300 & \nodata & \nodata & \nodata & \nodata \\
360 & \nodata & \nodata & \nodata & \nodata \\
\cutinhead{45$^\circ$ inclination}
180 &     5.0 &    11.1 &    13.5 &    14.8 \\
240 &     6.6 &    15.3 &    18.7 &    20.0 \\
300 & \nodata & \nodata &     7.1 &     9.9 \\
360 & \nodata & \nodata & \nodata & \nodata \\
\cutinhead{60$^\circ$ inclination}
180 & \nodata & \nodata & \nodata & \nodata \\
240 & \nodata &     8.0 &    11.2 &    13.2 \\
300 & \nodata &    10.2 &    14.0 &    15.5 \\
360 & \nodata &     8.8 &    14.1 &    17.1 \\
\enddata
\tablecomments{The values are averaged maximum velocities for all fibrils
               with that driver period and amplitude, corrected for
               projection effects. At the missing points,
               the movement of the transition region is irregular or has
               very low amplitude (<~5~km~s$^{-1}$).}
\end{deluxetable}

In Table~\ref{tbl:velocities}, we have listed the averaged maximum
velocities of all fibrils at given magnetic field inclinations, driver
amplitudes and periods. Unsurprisingly, at 0$^\circ$ inclination, only
180~s waves produce proper shocks, although it is possible to produce
regular movement of the transition region with 240~s waves if the driving
piston is strong enough. There is of course a dependence of the maximum
velocity on the driver amplitude, but it is not linear. The maximum 
velocity at 180~s period increases by only about 10\% when the driver
amplitude increases from 0.8 to 1.1~km~s$^{-1}$, a pattern that is 
repeated for several other combinations of period and inclination.
It also increases by only a factor of 2.4 when the amplitude increases
by a factor of 5.5, from 0.2 to 1.1~km~s$^{-1}$.
This indicates that the waves are reaching a plateau where larger
initial amplitudes just lead to increased dissipation.

The maximum velocity of 180~s waves decreases when the inclination increases.
Since these waves are well below the acoustic cutoff, this can not be
because they are becoming evanescent; instead, it is likely because of
the increased effective travel distance when the field is inclined. At
60$^\circ$ inclination, the effective distance to the transition region is
twice as long. This leads to increased dissipative losses as the wave
travels through the chromosphere. The effect is more pronounced at 180~s
because short-period waves steepen into shocks sooner and therefore are
more vulnerable to dissipation.

At the other periods, the maximum velocities increase as the acoustic
cutoff period increases, becoming noticeable when the wave period is
similar to the cutoff and significant when it is well below. At
60$^\circ$ inclination, the 240~s waves drop in amplitude again, while
those with longer periods are at their strongest.

It should be noted that field inclination is not the only possible
way to increase the cutoff period. We see in equation~\ref{eq:cutoff}
that we can also modify the sound speed, which is proportional
to the square root of the temperature. Hence, in a locally hotter
medium, we can get easier propagation of long-period waves. This
mechanism may be unlikely to be enough on its own, as you need
a temperature increase by a factor~2 to get the same cutoff increase
as a field inclination of 45$^\circ$, and a factor~4 to match 60$^\circ$;
however, the combination of the two mechanisms could be very effective.
For example, in the real solar atmosphere, the field tends to be
more vertical at lower heights but becomes more inclined higher up
due to the natural expansion of flux tubes. It is then possible that
a local temperature enhancement could increase the cutoff period in
the lower layers while the field inclination takes over as the waves
propagate upwards. Preliminary experiments using a simple model with
vertical field up to about 1000~km and inclined fields above that
indicate that such a configuration can also let long-period waves
propagate.

It could also be a possible reason why, in our simulations,
we see fewer fibrils with very high maximum velocities than
\citet{DePontieu+etal2007}. In our model, the field inclination
is constant, leading to a large increase in effective travel distance
at higher inclinations. In a more realistic model with the field
inclination increasing gradually, it is possible that the waves travel more
vertically through lower parts of the atmosphere before the field
inclination increases, leading to shorter travel distances and less
dissipation.

\section{Summary}

Our simulations have shown that, even with a simple 1D model, we are
able to reproduce the main observed properties of dynamic fibrils. We get
parabolic shapes and reproduce the range of decelerations and roughly the
range of maximum velocities found by \citet{DePontieu+etal2007}. We find
that the slope of the correlation between maximum velocity and deceleration
varies with the driver period, with lower decelerations at longer periods.
The distribution of decelerations is incompatible with a ballistic
model but fits very well with a shock wave deceleration model.
Furthermore, we have shown that long-period waves can propagate and
reach the transition region if they travel in a strong inclined field,
as suggested by \citet{Michalitsanos1973} and \citet{Bel+Leroy1977}, and
that they can drive fibrils there as suggested by \citet{Suematsu1990} and 
\citet{DePontieu+etal2004,DePontieu+etal2005}. The different slopes of
the correlation and the leakage of long-period waves
can account for the regional differences in the observations. Our
results give strong support to the theory that jets such as dynamic fibrils
and quiet sun mottles are driven
by slow mode magnetoacoustic shocks in the chromosphere.

\acknowledgements

This work was supported by the Research Council of Norway through grant
159483/V30. B.D.P. was supported by NASA grants NNG06-GG79G, NNG04-GC08G and
NAS5-38099 (TRACE) and would like to thank ITA for excellent hospitality in
August 2006.


\end{document}